\documentstyle[sprocl,epsfig]{article}

\def\lsim{\mathrel{\rlap{\lower4pt\hbox{\hskip1pt$\sim$}}
    \raise1pt\hbox{$<$}}}         
\def\gsim{\mathrel{\rlap{\lower4pt\hbox{\hskip1pt$\sim$}}
    \raise1pt\hbox{$>$}}}         

\def\be{\begin{equation}}
\def\ee{\end{equation}}
\def\bq{\begin{eqnarray}}
\def\eq{\end{eqnarray}}

\begin{document}

\title{SPIN AND TOTAL ANGULAR MOMENTUM OF THE GLUE}

\author{V. Barone}

\address{Dipartimento di
Fisica Teorica, Universit\`a di Torino, \\
 INFN, Sezione di
Torino, 10125 Torino, Italy\\
and  DSTA, Universit{\`a} `A. Avogadro',
15100  Alessandria, Italia \\
E-Mail: barone@to.infn.it}

\author{A. Drago}

\address{Dipartimento di Fisica, Universit{\`a} di Ferrara \\
and INFN, Sezione di Ferrara, 44100 Ferrara, Italy\\
E-Mail: drago@fe.infn.it}

\maketitle
\abstracts{
We briefly review the present theoretical 
and experimental knowledge on $\Delta G$ and 
$J_g$.}

\section{Introduction}

Polarized DIS \cite{mauro} 
probes the singlet axial charge of the proton $a_0$, 
i.e. the matrix element
\be
\langle P,S \vert \bar u \gamma_{\mu} \gamma_5 u + 
\bar d \gamma_{\mu} \gamma_5 d + \bar s \gamma_{\mu} \gamma_5 
s \vert P, S \rangle = 2 M \, a_0 \, S_{\mu}  
\label{1}
\ee
The surprising small value of $a_0$  ($\sim 0.2$)  
is explained as a consequence of the axial anomaly \cite{altarelli}
and of a relatively large  gluon contribution to the 
proton spin. 

Due to the axial anomaly $a_0$ does not coincide with 
the quark spin $\Delta \Sigma$ 
\be 
\Delta \Sigma = \int_0^1 dx \, (\Delta u + \Delta d + \Delta s)\;,
\label{2}
\ee
but mixes quark and gluon  
contributions in a scheme dependent way. 
In particular, in the Adler-Bardeen (AB) scheme $a_0$ 
is decomposed as 
\be
a_0(Q^2) = \Delta \Sigma -  N_f \, \frac{\alpha_s(Q^2)}{2 \pi} \,
\Delta G(Q^2) \;,
\label{3}
\ee
where $\Delta G$ is the gluon spin 
\be
\Delta G = \int_0^1 dx \, \Delta g \;.
\label{4}
\ee
In this scheme $\Delta \Sigma$ is conserved and it is natural to 
identify it with the total quark spin as provided by the constituent
quark model. 

Naively one has $\Delta \Sigma =1$. 
Due to relativistic effects
 and 
extra-degrees of freedom (e.g., mesons), 
 $\Delta \Sigma$
 is reduced to 0.6--0.7, still a moderately large value.  
According to (\ref{3}) the small  $a_0$ measured by 
the experiments is compatible with the
quark model estimate for $\Delta \Sigma$ 
  if one allows for a sizeable $\Delta G$. 

Two important questions are: 

\begin{itemize}

\item
Is it possible to infer 
the value of $\Delta G$ from  polarized 
DIS data ? 

\item
Do quark models, or  other nonperturbative approaches,  
provide a  $\Delta G$ compatible 
with the DIS findings ?

\end{itemize}

The answer to the first question is affirmative (although the result 
is affected by a large error) and comes from the global 
QCD analyses of the data. In Ref.~3 it has been 
shown that the DIS data suggest 
\be 
\Delta G = 1.4 \pm 0.9\;, \;\;\;\; 
{\rm at} \; \;Q^2 = 1\; \; {\rm GeV}^2\;, 
\label{deltagste}
\ee
where the error takes into account 
various sources of experimental 
and theoretical uncertainties (including different choices 
for the functional form of the helicity distributions).  

It is clear that a more direct measurement of $\Delta G$, 
 based on the investigation  of semiinclusive processes, 
is needed. 
A first result in this direction has recently come from 
 the HERMES study  of photoproduction  of high-$p_T$ hadron pairs
\cite{hermes}. 
The value of $\Delta g (x)/g(x)$ at one average $x$ point 
has been extracted from the longitudinal spin asymmetry by means 
of a Monte Carlo simulation. The result, referred to a scale 
$\mu^2 =2.1$ GeV$^2$, and obtained in LO QCD, is 
\be
\frac{\Delta g(x)}{g(x)} = 0.41 \pm 0.18 \,({\rm stat.}) \pm 
0.03 \, ({\rm exp}.\; {\rm syst.}) \;\;\;\;
{\rm at} \;\;  <x> =0.17
\label{deltaghermes}
\ee
Thus $\Delta g$ is found to be positive in 
the intermediate $x$ region. 
  The available LO parametrizations 
\cite{gehrmann,grsv} are in agreement with the value 
(\ref{deltaghermes}) (the exception being fit C of 
Gehrmann--Stirling \cite{gehrmann} where 
$\Delta g$ becomes negative above 
$x\simeq 0.1$ -- a behavior which seems to be ruled out by the 
HERMES finding). 
Although it is not possible, on the basis of the HERMES 
datum alone,  
 to discriminate among different fits and to draw  
definite conclusions about the integral of $\Delta g$, we can say
 that, if the distribution has
no bizarre behavior,  its first moment 
$\Delta G$ should also be positive.

As for the second question, the answer 
is affirmative as well. No discrepancies emerge between 
quark model expectations and results of 
DIS data analyses. 

Another interesting (and still open) problem is the value of the total 
angular momentum of the glue $J_g$, and of its orbital 
component $L_g = J_g - \Delta G$.  

All these issues will be briefly discussed in the following. 
To start with, let us first of all define  
 the quantities at hand.

\section{Angular momentum sum rule}

The angular momentum sum rule for the nucleon reads
\cite{ratcliffe,hoodbhoy}
\be
\frac 12=\frac 12 \, {{\Delta\Sigma}}+{{L_q}}(\mu^2)+
{{J_g}}(\mu^2)
\label{sumrule}
\ee
where $\Delta \Sigma$ is the quark helicity, eq.~(\ref{2}), 
$L_q$ is the quark canonical orbital 
 momentum and $J_g$  the total angular momentum of the 
glue. All quantities on the r.h.s. of eq.~(\ref{1}) are 
gauge invariant and depend in general on a scale $\mu^2$ 
(remember however 
that in the AB scheme  $\Delta \Sigma$ is a conserved 
quantity).  
The gluonic term in eq.~(\ref{sumrule}) is given by
\be
J_g(\mu^2) = \langle P \uparrow \vert 
\int \, d^3 r \, \left[\vec r \times \left ( \vec E (\vec r) \times
\vec B(\vec r) \right )\right]^3 \vert P \uparrow \rangle\, .
\label{jg}
\ee
$J_g$ admits a gauge invariant  decomposition
into an orbital ($L_g$) and a spin ($\Delta G$) part,
 but these two components cannot be written in terms of  
local operators. 
Adopting  the $A^+=0$ 
gauge, however, it is possible to obtain a local 
expression for  the gluon spin
\cite{manohar,jaffe}
\be
{{\Delta G}}(\mu^2)=\langle P\uparrow|\int{d^3 r \,
2 \, {\rm Tr}\left\{\left[{{\vec{E}(\vec{r})
\times\vec{A}(\vec{r})}}\right]^3+
{{\vec{A}_\perp(\vec{r})\cdot\vec{B}_\perp(\vec{r})}}\right\}}|P\uparrow
\rangle .
\label{deltag}
\ee
Both $J_g$ and $\Delta G$ have a scale 
dependence due to  
the renormalization of the operators 
 appearing in 
eqs.~(\ref{jg}, \ref{deltag}).  

Obviously,  $\Delta G$ and $J_g$ cannot be computed in 
perturbative QCD. However it is possible to evaluate 
them by means of nonperturbative tools. 
Calculations have been performed so far in the QCD sum rule framework 
\cite{balitsky,piller} or using  quark 
models \cite{jaffe,noi}. In the first case $J_g$ and $\Delta G$
are related to the quark and gluon condensates at a given
scale. In the second case 
one uses model wavefunctions to compute the 
matrix elements in eqs.~(\ref{jg}, \ref{deltag}). 
 The results are then assumed to be valid 
at some fixed scale, which is 
 usually very low ($\lsim 0.5$ 
GeV$^2$), and then evolved by the Altarelli-Parisi equations
to a higher $Q^2$.

\section{Sign and magnitude of $\Delta G$}

The sign of $\Delta G$ is an interesting matter.
Most of the existing fits yield a positive
$\Delta G$ at $Q^2 \ge 1$ GeV$^2$. 
 The nonperturbative results accumulated 
so far   also give positive values for $\Delta G$. 

The only indication for a negative $\Delta G$ came few years
ago from a model calculation by Jaffe \cite{jaffe}, who 
found $\Delta G \sim -0.4$ in the MIT 
bag model at a scale $\mu_0^2 \simeq 0.25$ GeV$^2$ and an 
even more negative value ($\sim -0.7$) in the non--relativistic 
quark model. It was later on explained  \cite{noi}
that Jaffe's negative result
is a consequence of neglecting self-interaction effects in the 
computation of (\ref{deltag}). 

To clarify this issue, let us see how (\ref{deltag}) 
reads in a quark model. 
 Denoting 
by $\langle \cdot 
\rangle_{\rm c}, \, \langle \cdot \rangle_{\rm osf}$ the color and 
orbital--spin--flavor expectation values, respectively,   
from (\ref{deltag}) one gets, after some manipulations 

\begin{eqnarray}
\Delta G &(\mu^2 )&=
\frac{1}{4} \, 
\sum_{{{i}},{{j}}=1}^3
\sum_{a=1}^{8} 
\langle  \lambda_i^a \lambda_j^a \rangle_{\rm c}   \nonumber \\
& \times &\int d^3r \, 
\langle P\uparrow|\Bigg\{
2\,  \left[\vec{E}_{{i}}({{\vec{r}}})\times
\vec{A}_{{j}}({{\vec{r}}})\right]^3+
\vec{A}_{\perp i}({{\vec{r}}})\cdot
\vec{B}_{\perp j}({{\vec{r}}})
\Bigg\}|P\uparrow\rangle_{\rm osf}\,.
\label{deltag2}
\end{eqnarray}
Now if one replaces the color fields 
by their expectation values in the nucleon eigenstate 
and includes the self-interaction terms with 
$i=j$,  $\Delta G$ turns out to be exactly zero. The reason 
is simple. Since the ground state of the nucleon is symmetric 
with respect to the exchange of any pair of quarks, the matrix elements
in the integral of 
eq.~(\ref{deltag2}) do not depend on the particle 
indices $i$ and $j$. Thus 
the integral 
factorizes out 
and $\Delta G$ vanishes exactly because 
$\sum_{i,j} \sum_a \langle \lambda_i^a  
 \lambda_j^a \rangle =0$. On the other hand,  if    
 the self--interaction terms 
are  neglected \cite{jaffe}, 
a negative $\Delta G$ is obtained,  
as a direct consequence of 
$ \sum_a \lambda^a_i \lambda^a_j = -8/3$ for $i \ne j$, 

As shown in Ref.~13, the correct way to proceed 
is to insert in the matrix element of (\ref{deltag2})
a complete set of intermediate states, which include
the orbital excitations of the three-quark system, and 
to take into account the self-interaction contributions. 

The integral in eq.~(\ref{deltag2}) 
no longer factorizes out
and $\Delta G$ is not forced to vanish, even including the 
self--field terms. 
We omit the details of this procedure (the reader can 
find them in Ref.~13). We just mention that, since we work 
in an effective model, which is supposed
to be valid up to excitation
energies of the order of 1
GeV, the convergence of the series obtained from the mode
expansion of (\ref{deltag2}) raises no problems: a cutoff 
is introduced, which excludes excited states   
with energies of more than  $\sim$ 1 GeV above the ground state.

In Ref.~13 we evaluated $\Delta G$ using the 
Isgur-Karl (IK) model \cite{IK}. We found 
\[
\Delta G = 0.26 \; \alpha_s \;. 
\]   
Here $\Delta G$ is expressed 
 in terms of the 
strong coupling constant $\alpha_s$, which is 
fixed in the IK model 
 by reproducing the $\Delta - N$ mass splitting. 
This gives $\alpha_s=0.9$ \cite{IK}, which  corresponds to
a scale $\mu_0^2 \simeq 0.25$ GeV$^2$. 
The dynamical quantities 
computed in the model must be
taken at this scale. 
Thus our final estimate is 
\be
\Delta G(\mu_0^2) \simeq 0.24\,,\;\;\;  
\mu_0^2 = 0.25 \, {\rm GeV}^2.
\label{deltagmu0}
\ee
The evaluation of the total angular momentum $J_g$ 
proceeds along  a similar line and leads to 
\be
J_g (\mu_0^2) \simeq 
\Delta G(\mu_0^2) \simeq 0.24\,,\;\;\;  
\mu_0^2 = 0.25 \, {\rm GeV}^2.
\label{jgmu0}
\ee
We conclude that at the model scale 
very little room (if any) is left 
for the orbital angular momentum of gluons: $L_g(\mu_0^2) 
= J_g(\mu_0^2) - \Delta G(\mu_0^2) \simeq 0$.

In order to evolve $\Delta G$ at NLO we need to know 
$\Delta \Sigma$. We cannot extract 
this quantity from the angular momentum sum rule unless we know  
$L_q$,  the canonical orbital momentum of quarks. 
We assume $L_q(\mu_0^2) = \pm 0.10$ (the available estimates  
\cite{hoodbhoy,scopetta} fall in this  range). 
With  (\ref{jgmu0}),  
eq.~(\ref{sumrule}) then implies $\Delta \Sigma = 0.52 \mp 0.20$. 

The NLO evolution of $\Delta G$ is shown in Fig.~1 where the 
shaded area corresponds to the allowed variation of  
$\Delta \Sigma$. We use $\Lambda = 250$ MeV 
for 3 flavors. Starting from $\Delta G = 0.24$ 
at $\mu_0^2$ we obtain, at 1 GeV$^2$,  
$\Delta G = 0.59 \pm 0.07$ which is compatible with 
the result (\ref{deltagste}) of the global analysis 
of Ref.~3 (the black dot with the error 
bar in figure). 

One may wonder whether a positive value of  $\Delta G$
at some  scale $Q^2 > 1$ GeV$^2$
is
compatible with a negative value at a smaller 
scale $\mu^2 \lsim 0.5$ GeV$^2$ (a typical 
quark model scale). In other terms, is it possible
to obtain by QCD evolution a positive  $\Delta G$
starting from a negative input at low energy~?
In principle, the answer is yes. In practice, however, 
a $\Delta G$  such as that demanded by the DIS data 
implies a positive gluon polarization even at very 
small scales, as we shall now see.

\begin{figure}[tbp]
\vspace{3truecm}

\parbox{6cm}{
\epsfig{figure=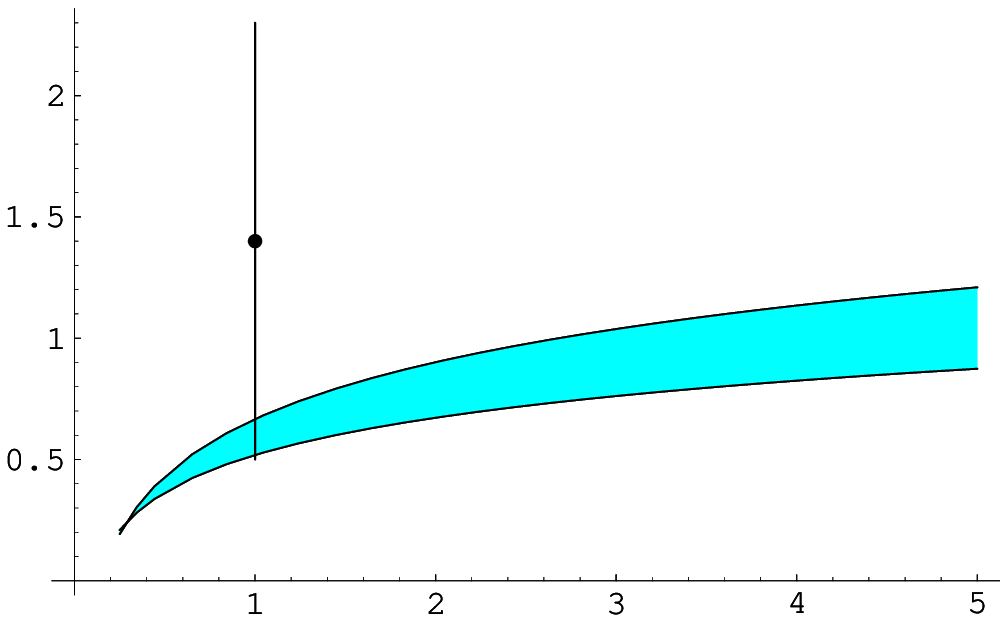,width=6cm}
\hspace*{1truecm}
\parbox{4.5cm}{\small Figure 1: 
$\Delta G$ as a function of $Q^2$ (in GeV$^2$) 
in NLO QCD. The input
 is $\Delta G = 0.24$ at $\mu_0^2 =0.25$ GeV$^2$. 
The black dot with error bar is the estimate of 
Ref.~3, based on a global analysis of DIS data.}
}
\nolinebreak
\parbox{6cm}{
\epsfig{figure=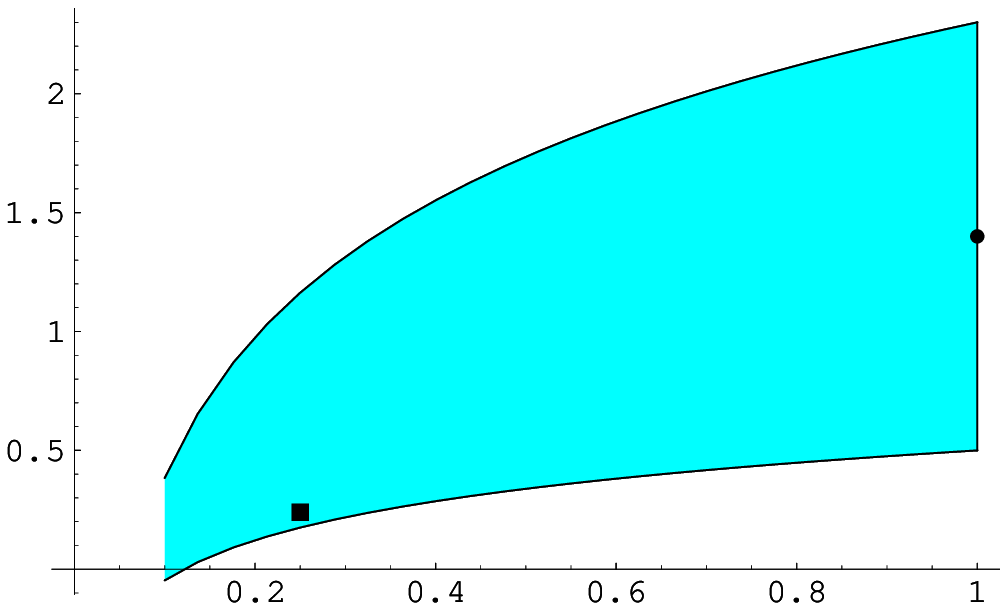,width=6cm}
\hspace*{1truecm}\parbox{4.5cm}
{\small Figure 2: 
Backward evolution of $\Delta G$ (in abscissas
$Q^2$ expressed in GeV$^2$). The starting value is~\cite{stefano} 
 $\Delta G = 1.4 \pm 0.9$ at 1 GeV$^2$ 
(black circle with error). The squared dot is the result of the 
model computation of  Ref.~13.}
}
\vspace{0.5truecm}
\end{figure}

Let us take a look at the NLO evolution for  $\Delta G$
in the AB scheme: 
\begin{eqnarray}
\Delta G(Q^2) =
&&\left [1 - \frac{2 \, N_f}{\pi \, \beta_0} \left 
( \alpha_s(\mu^2) - \alpha_s(Q^2) \right ) \right ]\, 
\frac{\alpha_s(\mu^2)}{\alpha_s(Q^2)} \, \Delta G(\mu^2)   \nonumber \\
&-&\frac{4}{\beta_0}\, \left (1 - \frac{\alpha_s(\mu^2)}{\alpha_s(Q^2)}
\right ) \, \Delta \Sigma \;, 
\label{deltagQCD}
\end{eqnarray}
with $\beta_0 = 11 - 2 N_f /3$. 
It is clear that the second term of (\ref{deltagQCD}) 
has no definite sign, so nothing prevents 
a negative $\Delta G$ at low $Q^2$ from producing a positive 
one at high $Q^2$. But if we evolve back 
the value (\ref{deltagste}) we find that $\Delta G$ 
remains positive down to very small $Q^2$ values 
($\sim 0.15-0.20$ GeV$^2$), below which the perturbative
evolution equations are simply inapplicable. 
This is shown in Fig.~2 where the backward evolution 
of $\Delta G$ is plotted. The shaded area corresponds to the error 
in (\ref{deltagste}). For $\Delta \Sigma$ we used the 
estimate of Ref.~3: $\Delta \Sigma = 0.44$. The uncertainty  
quoted on this value ($\pm 0.09$) does not produce any 
appreciable variation on the final result. 

\section{The total angular momentum of the glue}

An open issue \cite{balitsky} is the relative weight 
of $\Delta G$ and $J_g$, the gluon total angular momentum, or
-- equivalently -- the sign and the size of the gluon 
orbital angular momentum $L_g$. 

We have already mentioned that  the 
 calculation of eq.~(\ref{jg}) in the 
IK model \cite{noi} gives $
J_g (\mu_0^2) \simeq 
\Delta G(\mu_0^2) \simeq 0.24$ and hence  
$L_g (\mu_0^2) \simeq 0$. 

To see what happens at a larger scale
we perform a   
leading--order  QCD evolution of $J_g$ \cite{hoodbhoy}
(a semiquantitative result is enough for our 
purposes). 
The result (with $\Lambda = 200$ MeV for 3 flavors) is 
\be
J_g \, (1 \; {\rm GeV}^2) = 0.29\;. 
\label{jg1}
\ee
A QCD sum rule calculation \cite{balitsky} gives a similar value 
\be
J_g (1 \; {\rm GeV}^2) = 0.35 \pm 0.13\;.  
\ee

Note that $\Delta G$ evolves with $Q^2$ differently 
from $J_g$. At LO, for instance, whereas $\Delta G$ increases  
as $\ln {(Q^2/\Lambda^2)} / \ln {(Q_0^2/\Lambda^2)}$, 
 $J_g$  has a limiting asymptotic value
 $8 / (16 + 3 N_f) 
= 0.32$. 

Hence, with growing $Q^2$, $\Delta G$ gets larger 
than $J_g$ and 
the gluon orbital angular 
momentum $L_g$ 
becomes increasingly negative.  This scenario is 
similar to the one suggested by the QCD sum rule approach  
of Ref.~11 and by the quark model 
calculation of Ref.~15. 

What remains to be done, in the framework 
of quark models, is a check of the 
angular momentum sum rule, eq.~(\ref{sumrule}), 
with a self-consistent evaluation of all 
terms appearing in it. This work is in progress.

\section{Summary}

In conclusion, let us summarize the present information 
on $\Delta G$ and $J_g$. 

\begin{itemize}

\item
Inclusive DIS data require \cite{stefano}
\be
\Delta G \, (1 \, {\rm GeV}^2) = 1.4 \pm 0.9\,, 
\label{forte}
\ee
where 
the error takes into account various  
uncertainties.  

Evolving this value back in $Q^2$ one still gets 
a positive $\Delta G$ down to $Q^2 \simeq 0.15-0.20$ 
GeV$^2$, which is the limit of applicability of 
the evolution equations.

\item
A recent semiinclusive measurement gives at a scale 
$\mu^2 = 2.1$ GeV$^2$
\be
\frac{\Delta g}{g} = 0.41 \pm 0.18\, ({\rm stat.}) 
\pm 0.03 \, ({\rm exp.} \; {\rm syst.})\;, 
\;\;\;\; {\rm at} \;\; <x> =0.17 \;.
\ee
This hints to a positive $\Delta G$. 
More precise and extensive determinations are needed in order 
to draw any definite conclusion.

\item
Contrary to some claims, 
quark models do not prescribe a negative gluon helicity: this 
is only an artifact of neglecting self-interaction 
contributions. If these are included, which is the correct
thing to do, a positive $\Delta G$ results at the model scale. 
In particular, in the IK model, we found  
\[ 
\Delta G = 0.24
\]
 Setting  $\Delta \Sigma = 0.52 \pm 0.20$,  $\Delta G$ becomes 
at 1 GeV$^2$
\[
\Delta G = 0.59 \pm 0.07 \;,
\]
to be compared with (\ref{forte}). The two values are compatible 
within the errors.

\item
Quark model and QCD sum rule calculations indicate a 
small gluon orbital contribution at low $Q^2$. 
 In the IK model, for instance, we obtained 
\[
\Delta G \simeq J_g\;, \;\;\; {\rm i.e.} \;\;\; L_g \simeq 0 
\;\;\; {\rm at} \;\; \mu_0^2 = 0.25 \; {\rm GeV}^2   
\]
which implies at 1 GeV$^2$ 
\[
J_g = 0.29 \;. 
\]
A close result was found in the QCD sum rule 
calculation of Ref.~11.
Since $J_g$ evolves slower than $\Delta G$ 
 we expect 
the gluon orbital momentum to be increasingly negative
as $Q^2$ gets larger.

\end{itemize}


\section*{References}
\begin{thebibliography}{99}


\bibitem{mauro}
Two excellent reviews on the subject are: \\ 
M.~Anselmino, A.~Efremov, E.~Leader, 
Phys. Rep. 261 (1995) 1. \\
B.~Lampe and E.~Reya, {\tt hep-ph/9810270}. 



\bibitem{altarelli}
G.~Altarelli and G.G.~Ross, Phys. Lett. B 212 (1988) 391. 
A.V.~Efremov and O.V.~Teryaev, JINR report E2-88-287 (1988). 
R.D.~Carlitz, J.C.~Collins and A.H.~Mueller, Phys. Lett. 
B 214 (1988) 229.


\bibitem{stefano}
G.~Altarelli, R.D.~Ball, S.~Forte and G.~Ridolfi, Nucl. Phys. 
B 496 (1997) 337; {\tt hep-ph/9803237}

\bibitem{hermes}
P.K.A.~ de Witt Huberts, these proceedings. M.~Amarian, 
contribution to DIS99, Zeuthen. 

\bibitem{gehrmann}
T.~Gehrmann and W.J.~Stirling, Phys. Rev. D 53 (1996) 6100. 

\bibitem{grsv}
M.~Gl{\"u}ck, E.~Reya, M.~Stratmann and W.~Vogelsang, 
Phys. Rev. D 53 (1996) 4775. 

\bibitem{ratcliffe}
P.G.~Ratcliffe, Phys. Lett. B 192 (1987) 180.

\bibitem{hoodbhoy} 
X.~Ji, J.~Tang and P.~Hoodbhoy, Phys. Rev. Lett. 76 (1996) 740.
X.~Ji, Phys. Rev. Lett. 78 (1997) 610. 


\bibitem{manohar}
A.~Manohar, Phys. Rev. Lett. 65 (1990) 2511; 66 (1991) 289. 


\bibitem{jaffe}
R.L.~Jaffe, Phys. Lett. B 365 (1996) 359.

\bibitem{balitsky}
I.~Balitsky and X.~Ji, Phys. Rev. Lett. 79 (1997) 1225. 


\bibitem{piller}
L.~Mankiewicz, G.~Piller, A.~Saalfeld, 
Phys. Lett. B 395 (1997) 318.

\bibitem{noi}
V.~Barone, T.~Calarco and A.~Drago, 
Phys. Lett. B 431 (1998) 405. 


\bibitem{IK}
N.~Isgur and G.~Karl, Phys. Rev. D 18 (1978) 4187; 
Phys. Rev. D 19 (1979) 2653. M.~Brack and R.K.~Bhaduri, 
Phys. Rev. D 35 (1987) 3451. 



\bibitem{scopetta}
S.~Scopetta and V.~Vento, {\tt hep-ph/9901324}. 








\end{thebibliography}
\end{document}